\documentclass[10pt,twocolumn,twoside]{IEEEtran}

\usepackage{graphicx,epsfig}
\usepackage{cite}
\usepackage{stfloats}
\usepackage{amsmath}
\usepackage{amsfonts,helvet}

\newtheorem{theorem}{Theorem}

\newtheorem{algo}[theorem]{Algorithm}

\newtheorem{proposition}[theorem]{Proposition}

\newcounter{proposition}
\begin{document}
\title{Novel Modulation Techniques using Isomers as Messenger Molecules for Nano Communication Networks via Diffusion}


\author{\authorblockN{Na-Rae Kim,~\IEEEmembership{Student~Member,~IEEE} and Chan-Byoung Chae, \IEEEmembership{Senior~Member,~IEEE}}
\thanks{The authors are with the School of Integrated Technology, Yonsei University, Korea. Email: \{nrkim, cbchae\}@yonsei.ac.kr.}
\thanks{This work was in part supported by the Ministry of Knowledge Economy under the ``IT Consilience Creative Program"  (NIPA-2012-H0201-12-1001) and the Yonsei University Research Fund of 2011. Part of this work was presented at the \emph{IEEE International Conference on Communications (ICC workshop on Molecular and Nano-Scale Communications)}, June, 2012~\cite{Chae_ICC12}.}}




\maketitle \setcounter{page}{1} 
%
%
%

\markboth{submitted for journal publication}%
{Kim and Chae: Nano Communication Networks via Diffusion}

\begin{abstract}
In this paper, we propose three novel modulation techniques, i.e., concentration-based, molecular-type-based, and molecular-ratio-based, using isomers as messenger molecules for nano communication networks via diffusion. 
To evaluate achievable rate performance, we compare the proposed techniques with conventional insulin based concepts under practical scenarios. Analytical and numerical results confirm that the proposed modulation techniques using isomers achieve higher data transmission rate performance (max~7.5~dB signal-to-noise ratio gain) than the insulin based concepts. We also investigate the tradeoff between messenger sizes and modulation orders and provide guidelines for selecting from among several possible candidates. 
\end{abstract}

\begin{keywords}
Nano communication networks, molecular communication, modulation technique, isomer, diffusion, and messenger molecule.
\end{keywords}


\section{Introduction}
\PARstart{S}{ince} Richard Feynman's talk on top-down nanotechnology, people have eagerly pursued practical work on smaller and smaller scales, most notably through nanotechnology~\cite{Feynman}. Nanotechnology, recently, produced a new branch of research called nano communication networks (NCNs)~\cite{Akyildiz_cn08,JSAC_10,Akyildiz_cn09}. NCNs interconnect several nanoscale machines (a.k.a. nanomachines) to perform simple tasks or to carry out more complex tasks in a cooperative manner~\cite{Akyildiz_cn08,JSAC_10}. These networks are not just smaller versions of traditional communication networks; they have their own features and are applicable in many fields, including biomedical, industrial, military, and environmental~\cite{Akyildiz_cn09}.

NCNs can be realized by several methods.
We can rely on traditional communication systems that use electromagnetic fields or ultrasonic waves. Such a method, however, has to overcome some radio frequency (RF) device barriers~\cite{Akyildiz_cn08}. On the practical front, researchers are considering new materials, such as carbon-nano tubes (CNTs) and graphene~\cite{Akan_commmag10,Akyildiz_mag10}. Areas that have yet to receive much research, however, are channel models and human body absorption with terahertz bands. Therefore, researchers have introduced a new concept utilizing diffusion that is especially useful for short range communication~\cite{short_diffusion}. 



This new paradigm of communication, molecular communication, sends/receives information-encoded molecules between nanomachines~\cite{Akyildiz_cn08,Hiyama_mc,Pierobon_EuCAP10,Gregori_cn10,Moore_05}. Several papers have addressed the achievable transmission rates (achievable rate hereafter) of the communication system, and the author in~\cite{JSAC_10} evaluated, using a circuit model, the normalized gain and delay of the system. Though a great deal of these papers have ignored possible noise sources, recently several works have focused on noise analysis~\cite{TSP_11,collision,Moore_NBS09}. 

One of the advantages of molecular communication is its biocompatibility. By using bio-molecules, we can create inherently biocompatible systems that require no harmful inorganic materials. Indeed, biocompatibility
is regarded by many as the most important and challenging issue concerning intra-body applications. Molecular communication, driven by chemical reactions, is also energy efficient~\cite{Moore_05}. For these reasons, this paper mainly focuses on molecular communication. 

The authors in~\cite{Kuran_NCN10,Kuran_ICC11} studied extensively the fundamentals of molecular communication via diffusion. In \cite{Kuran_NCN10}, they investigated a new energy model to understand how much energy is required to transmit messenger molecules and \cite{Kuran_ICC11} introduced concentration-based and molecular-type-based modulation techniques. The researchers also have compared, by using a simple binary symmetric channel model, the achievable rate. In analyzing their modulation techniques, however,~\cite{Kuran_NCN10,Kuran_ICC11} did not clearly suggest concrete structures for the messenger molecules. They considered insulin-based nano networks, but it is unclear how these could be utilized in practice. In this paper, so as to maximize the achievable rate with less transmit power/energy, we propose using isomers as messenger molecules. We also suggest a new ratio-based modulation method. To consider the properties of isomers, we slightly modify the energy model described in~\cite{Kuran_NCN10}. To the best of our knowledge, the proposed method is the first attempt to design appropriate practical messenger molecules for NCNs via diffusion.

This paper is organized as follows. Section~\ref{Sec:Main} describes the channel and energy model under consideration. Section~\ref{Sec:ModTech} explains isomers for messenger molecules and also proposes three modulation techniques, i) concentration-based, ii) molecular-type-based, and iii) ratio-based. We present the numerical results in Section~\ref{Sec:Num}, and Section~\ref{Sec:Conc} offers our conclusions.

\begin{figure}[t]
 \centerline{\resizebox{1\columnwidth}{!}{\includegraphics{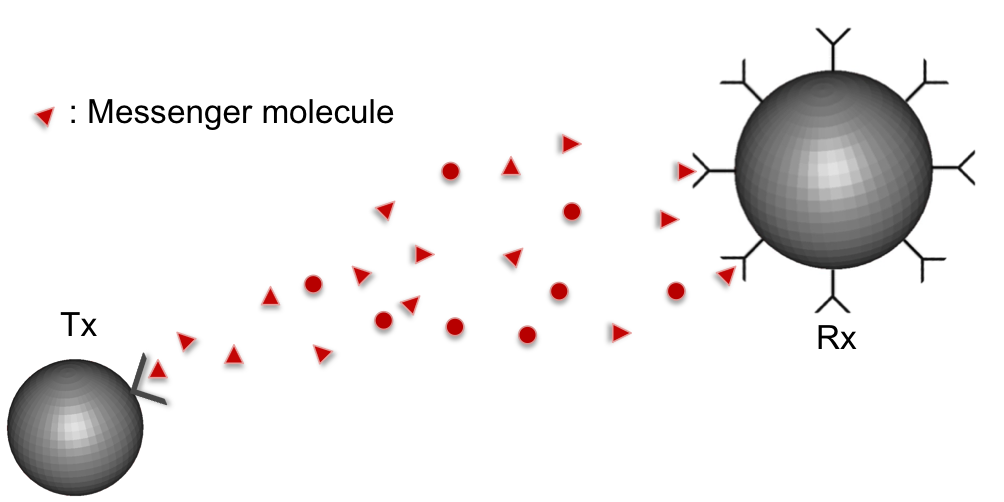}}}
  \caption{System model of nano communication via diffusion with a single transmitter and a single receiver.}
  \label{Fig:SysModel}
\end{figure}

\section{System Model}
\label{Sec:Main} 
We consider a nano communication system that, for simplicity, consists of a single transmitter and a single receiver, as illustrated in Fig.~\ref{Fig:SysModel}. The information-encoded molecules are called messenger molecules, and they can propagate through the medium by several methods. Three main NCN architectures are based on the propagation type (e.g., walkaway-based, flow-based, and diffusion-based)~\cite{JSAC_10}. For this system, the diffusion-based propagation is chosen for analysis, so the messenger molecules diffuse through the medium (i.e., liquid) at body temperature. This paper assumes that no collisions occur among the propagating messenger molecules.

\subsection{Channel Model}
\label{Subsec: Channel}
\begin{figure}[t]
 \centerline{\resizebox{1\columnwidth}{!}{\includegraphics{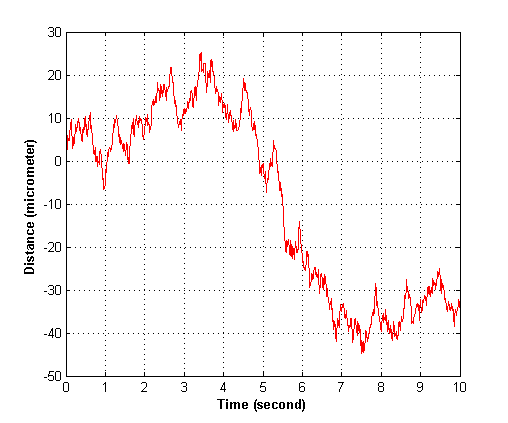}}}
  \caption{An example of the Brownian motion of a particle in 1-$D$ motion.}
  \label{Fig:Brownian}
\end{figure}

\begin{figure}[t]
 \centerline{\resizebox{1\columnwidth}{!}{\includegraphics{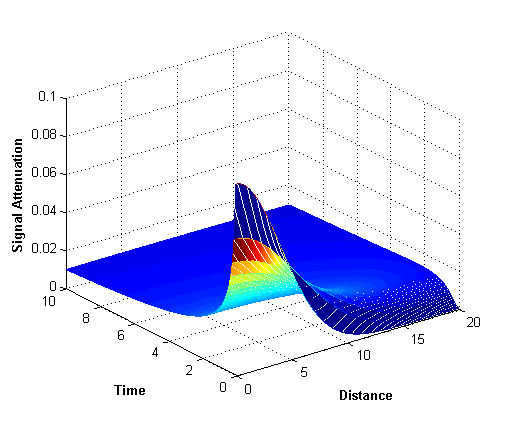}}}
  \caption{Signal attenuation as functions of distance and time.}
  \label{Fig:SigAttn}
\end{figure}

The particles or messenger molecules released from a transmitter nanomachine spread out through the medium by Brownian motion~\cite{Kuran_NCN10,TSP_11}. Such motion is basically driven by diffusion, meaning the particles move from areas of higher concentration to areas of lower concentration, and the displacement of messenger molecules follows a normal distribution with zero mean. Fig.~\ref{Fig:Brownian} is one example of one-dimensional Brownian motion. Also, Fig.~\ref{Fig:SigAttn} shows molecular signal attenuation as functions of distance and time. The standard deviation, $\sigma$, of the displacement can be obtained as follows:
\begin{align}
&\bigtriangleup{X} \sim \mathcal{N}(0, \sigma^2), \nonumber \\
&\frac{\partial{c}}{\partial{t}}=D\frac{{\partial^2}c}{\partial{x^2}},\label{Fick's}\\
&c(x,t)=\frac{1}{(4{\pi}Dt)^{1/2}}e^{-x^2/{4Dt}},\label{c}\\
&\overline{x^2} = 2Dt,\nonumber\\
&\sigma=\sqrt{2Dt},\nonumber\\
& D = \frac{K_b T}{6\pi \eta r_{mm}} \label{D}
\end{align}
where $c$ indicates the concentration of Brownian particles at time $t$ at point $x$, and $D$ represents the diffusion coefficient of the particles calculated from the following: Boltzmann constant ($K_b$), the temperature ($T$), the viscosity of the medium ($\eta$), and the radius of a messenger molecule ($r_{mm}$). Eq. (\ref{Fick's}) is Fick's second law of diffusion. By solving this partial differential equation, we obtain the general equation for the concentration of particles represented in (\ref{c}). The first moment of (\ref{c}) is zero, which indicates the displacement has a zero mean, and the second moment becomes the variance. Hence, the displacement has a standard deviation of $\sqrt{2Dt}$, and finally we have
\begin{align}
\bigtriangleup{X} \sim \mathcal{N}(0, 2Dt).
\end{align}

When $n$ number of messenger molecules are transmitted by the transmitter nanomachine, the molecules have a probability of hitting the receiver nanomachine. We represent this as a binomial distribution of $n$ times of trials with a hitting probability ($P_\text{hit}$) for each trial. $P_\text{hit}$ is determined by the symbol duration ($T_s$) and the distance between the transmitter and the receiver ($d$), which are both affected by the diffusion coefficient ($D$). If $n$ is large enough, and $nP_\text{hit}$ is not zero, the binomial can be approximated as a normal distribution. In addition, we have to take into account the overflow molecules from the previous symbol as well as the molecules sent from the current symbol.\footnote{In our paper, we consider the overflow only from the previous symbol since $P_\text{hit}$ for $2T_s$ shows a significant increase compared to $P_\text{hit}$ for $T_s$, but, as more time passes, no meaningful difference exists(i.e., memory channel)~\cite{Kuran_NCN10}.}
Thus, the total number of molecules during one symbol duration differs according to the current and previous symbols. The following represents the number of molecules from the current and previous symbols and the noise term.
\begin{align}\begin{split}
&Binomial (n, P_\text{hit}(d,T_s))\\
&\sim \mathcal{N}(nP_\text{hit}(d,T_s), nP_\text{hit}(d,T_s)(1-P_\text{hit}(d,T_s))),\\
&N_c \sim \mathcal{N}(np_{1}, np_{1}(1-p_{1})),\\
&N_p \sim \mathcal{N}(np_2,np_2(1-p_2))-\mathcal{N}(np_1,np_1(1-p_1)),
\\
&N_n \sim \mathcal{N}(0, \sigma^2), \end{split}
\label{normal}  \end{align}
where, $N_c$ denotes the number of molecules transmitted and received during the current symbol duration, and $N_{p}$ denotes the number of molecules transmitted in the previous symbol duration but received during the current symbol duration. For simplicity, we use $p_1$ for $P_\text{hit}$ during $T_s$ and $p_2$ for $P_\text{hit}$ during $2T_s$. The molecules from noise sources can be summed up as $N_n$, and, in this paper, it is assumed as Additive White Gaussian Noise (AWGN)~\cite{Kuran_ICC11}.


\subsection{Energy Model}
\begin{figure}[t]
 \centerline{\resizebox{1.1\columnwidth}{!}{\includegraphics{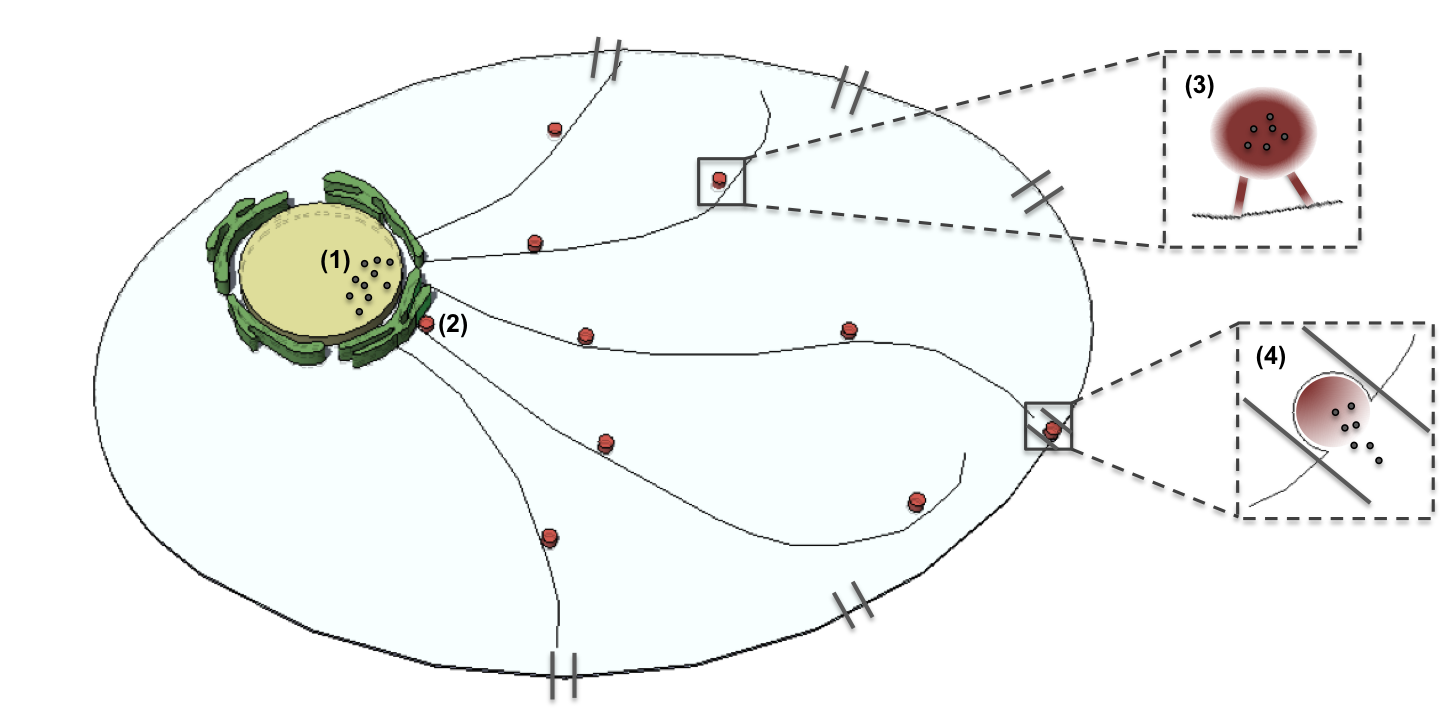}}}
  \caption{Energy model of nano communication composed of four steps. The first step (1) is for generating the messenger molecules, (2) for synthesizing the vesicles, (3) is for carrying them, and (4) is for extracting them to the medium.}
  \label{Fig:EneModel}
\end{figure}

We use the energy model described in~\cite{Kuran_NCN10} and assume that basic eukaryotic cells are a perfect model for nanomachines. First, the messenger molecules are synthesized inside a nucleus and encapsulated in vesicles during intra-cellular propagation. The vesicles are then carried to the boundary of the transmitter nanomachine. Lastly, the messenger molecules are released into the propagation medium through membrane fusion of vesicles. Each step is illustrated in Fig.~\ref{Fig:EneModel}. Fig.~\ref{Fig:EneModel} represents a simplified form of an eukaryotic cell structure to model a nanomachine. The large circle represents the cell membrane filled with the medium (e.g., cytoplasm), and the inner circle represents the nucleus of a cell. The structure around the nucleus is a Golgi apparatus that produces and transports vesicles. Furthermore, some interconnections exist between the Golgi and the cell boundary--microtubules or molecular rails--which messenger molecule-carrying vesicles travel along. In the cell membrane, several gated channels exist that the nanomachine can, as needed, control the on/off of them. Specifically, the gated ionic channel can be on/off depending on the presence of ions near the channel.

The energy needed in each step is calculated. Moreover, we can obtain the cost of synthesizing messenger molecules (hexoses as an example) by the enthalpy of formation ($\Delta H$, which is the change of enthalpy that accompanies the formation of one mole of the compound from its elements). The final energy model can then be presented as follows: (see~\cite{Kuran_NCN10} for more details)
\begin{align}
E_T& = nE_S+\frac{n}{c_v}(E_V+E_C+E_E),\nonumber\\
E_S&= \frac{\Delta H_{hexose}}{6.02\times10^{23}}  J\text{ per messenger molecule},\nonumber
\end{align}
\begin{align}
E_V&=83\times5(4\pi{r_v}^2) zJ,\nonumber\\
E_C&=83\times \left[\frac{r_{unit}/2}{8}\right]zJ,\nonumber \\
E_E&=83\times10 zJ,\nonumber\\
c_v&={\left(\frac{r_v}{r_{mm}\sqrt{3}}\right)}^3,\nonumber\\
zJ&={10}^{-21}J\nonumber
\end{align}
where, $E_T$ is the total energy cost required to transmit $n$ number of molecules, $E_S$ is the synthesizing cost of one hexose molecule calculated from the sum of bond energies (e.g., the enthalpy change), and $E_V$ is the vesicle-synthesizing cost having a radius of $r_v$. $E_C$ is the cost of intra-cellular transportation having cell radius of $r_{unit}$, $E_E$ is for membrane fusion, and $c_v$ is the vesicle capacity, the number of messenger molecules one vesicle can carry, which is related to the radius of messenger molecules $r_{mm}$. 

\section{Modulation Techniques}
\label{Sec:ModTech}
In molecular communication via diffusion, the unique properties of the messenger molecules can determine modulation techniques. Two such techniques proposed in~\cite{Kuran_ICC11}  include the use of concentration and types of messenger molecules. A usable messenger molecule itself, however, has yet to be proposed. Thus, this paper proposes practical messenger molecules as well as a new modulation technique using ratios of the molecules. We also analyze and compare achievable rates by applying different modulation techniques.

  \begin{figure*}[!t]
 \centerline{\resizebox{1.9\columnwidth}{!}{\includegraphics{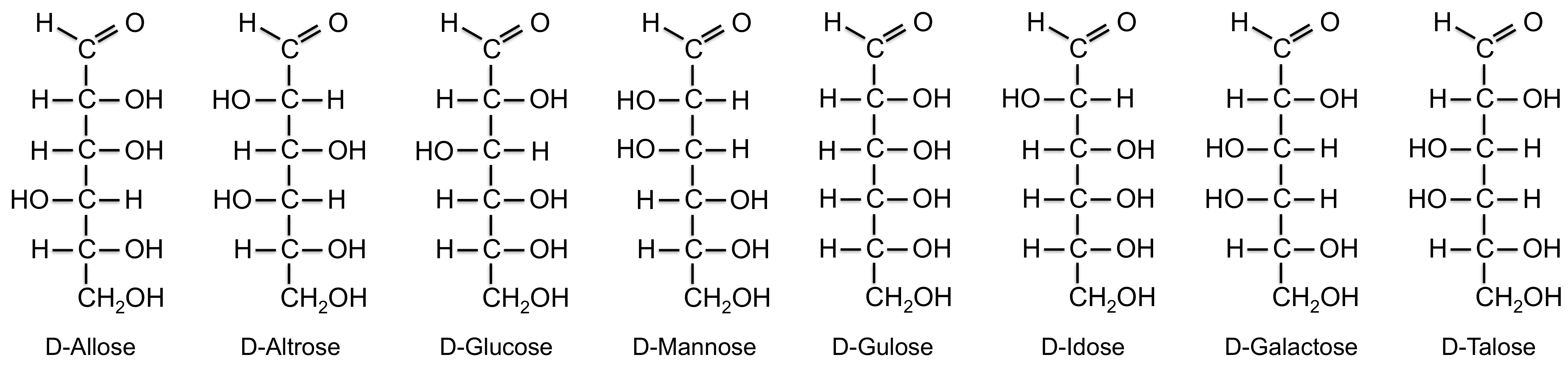}}}
  \caption{$D$-form isomers of aldohexoses.}
  \label{Fig:hexose}
\end{figure*}

\subsection{Isomers for Messenger Molecules}  
\label{SubSec:Isomer}
The most important thing when designing messenger molecules is that they have to be non-toxic to the human body. The one suggested in \cite{Kuran_NCN10}, however, is highly flammable (i.e., hydrofluorocarbon), which means it may be inappropriate for practical applications. 

For several reasons, potential candidates for messenger molecules are isomers, molecules composed of the same number and types of atoms \cite{BIOCHEM}. First of all, they consist of the same type of atoms, lightening the burden at the transmitter nanomachine that synthesizes them. For numerical analysis, this paper uses the isomers known as hexoses, especially aldohexoses (i.e., aldose forms of the hexoses). Note that we can select one from the aldoses family (e.g., hexoses, pentoses, tetroses, or trioses) based on the required modulation order.

\begin{figure}[!t]
 \centerline{\resizebox{0.53\columnwidth}{!}{\includegraphics{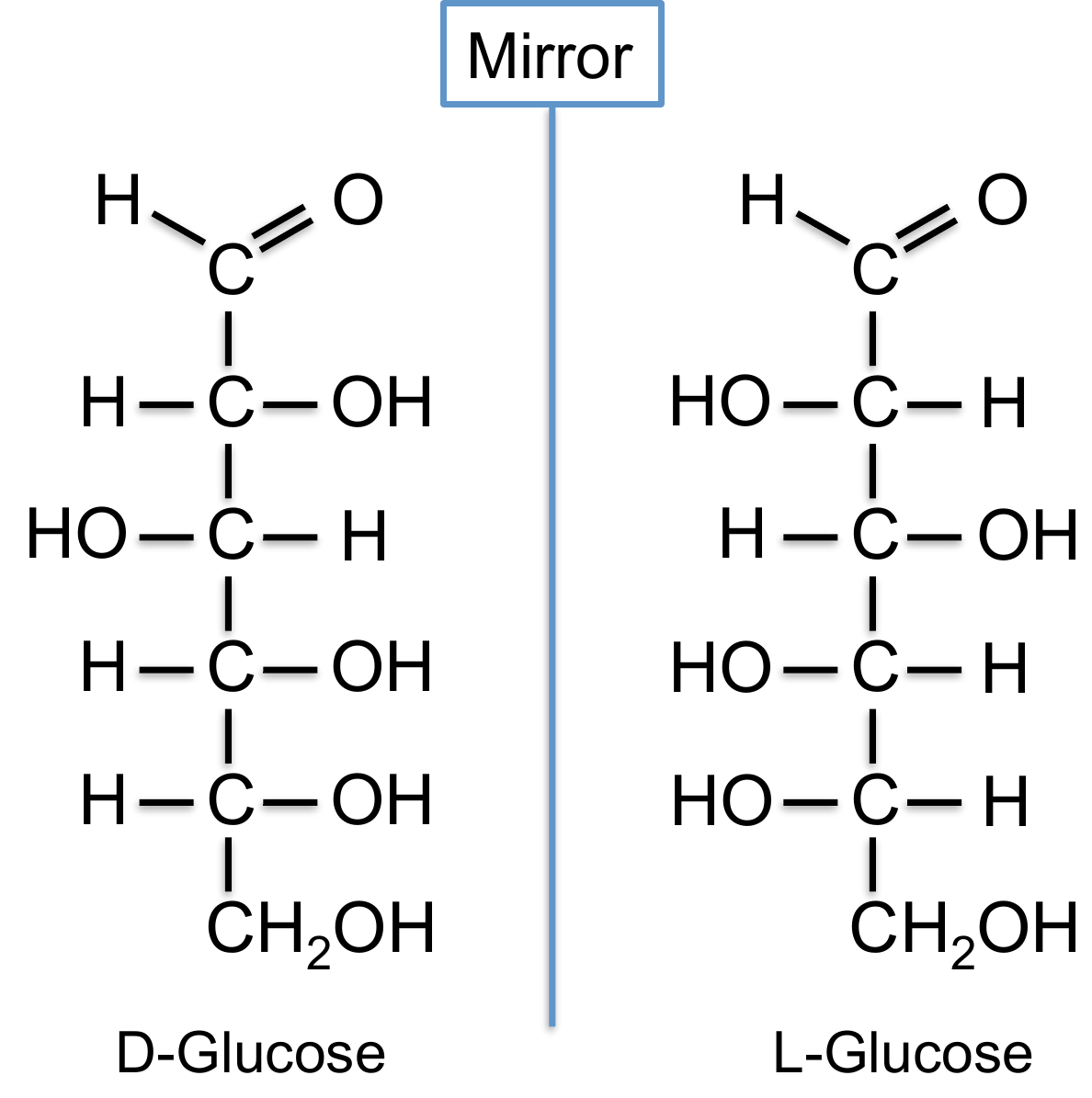}}}
  \caption{Mirror image of $D$- and $L$- glucose. They are enantiomers.}
  \label{Fig:mirror}
\end{figure}

Aldohexoses are monosaccharides with the chemical formula of  $C_6H_{12}O_6$. They have four chiral carbon atoms, not superposable on the mirror images, which give them $16$ (=$2^4$) stereoisomers~\cite{chemterm}. Fig.~\ref{Fig:hexose} illustrates eight kinds of $D$- form diastereomers. The enantiomers of each molecule are another set of eight $L$- form diastereomers. Therefore, aldohexoses have 16 different shapes. Here, diastereomers are stereoisomers that are not enantiomers (mirror-image isomers). For example, $D$-glucose and $L$-glucose as shown in Fig.~\ref{Fig:mirror} are enantiomers. On the other hand, $D$-glucose and $D$-galactose--isomers but not mirror images--are diastereomers. 

\begin{figure}
 \centerline{\resizebox{1\columnwidth}{!}{\includegraphics{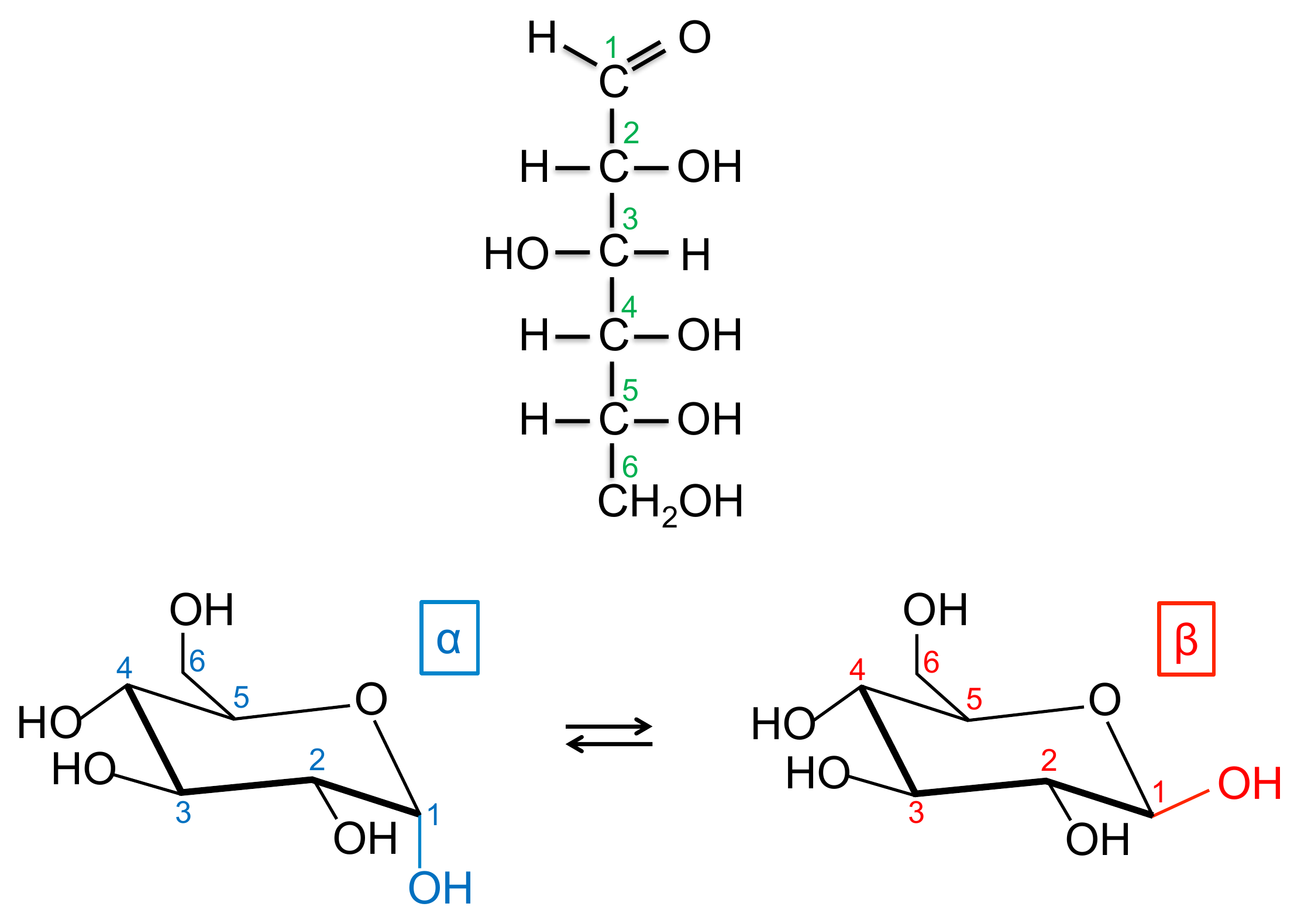}}}
  \caption{$D$-glucose and its $\alpha$- and $\beta$- anomers. They undergo rapid interconversion in water, which is called mutarotation.}
  \label{Fig:muta}
\end{figure}

When each isomer is dissolved in an aqueous solution, it mostly exists as a cyclic form (i.e., a ring form). $D$-glucose, for instance, undergoes nucleophilic addition reaction generating four cyclic anomers, $\alpha$-, $\beta$- forms of $D$-~glucopyranose and $D$-glucofuranose~\cite{nucleo}. We consider, however, only two pyranose ($\alpha$-, $\beta$-) forms since they predominate in the solution with $36$ and $64$ percentages, respectively. If the functional groups attached to carbon number~1 ($C1$) and $C5$ shown in Fig.~\ref{Fig:muta} have a trans-structure, it is called $\alpha$- form, and if a cis-structure, $\beta$- form. They interconvert each other in solution through a process called mutarotation. Therefore, by deploying hexoses groups into the system, there are 32 different isomers in total. 

\subsection{Molecular-Concentration-Based}
\label{Subsec:CSK}
When the concentration of messenger molecules is used to differentiate symbols, the technique is known as concentration-based modulation, originally introduced in~\cite{Kuran_ICC11}. Different concentration values represent different symbols, which are divided by proper thresholds. This is called concentration shift keying (CSK), and also can be referred to as isomer-based CSK (ICSK) here since we use isomers as the messenger molecules. In binary-ICSK (B-ICSK), one threshold exists to represent two different symbols, and a receiver nanomachine decodes the symbol as `1' if the number of the received messenger molecules exceeds the threshold ($\tau$), `0' otherwise. Generally, 2$^n$-ICSK systems transmit $n$ bits per symbol, and requires 2$^n$-1 number of thresholds. In this system, there is, theoretically, no limit in the modulation order. As the modulation order increases, however, so does the error probability since the minimum distance between two neighboring thresholds decreases. The system only uses one kind of molecule, and $D$-glucopyranose is used for analysis here. 
\subsubsection{B-ICSK}

\begin{align}\begin{split}
&P_a(0,0) = P_b(0,0,0) + P_b(1,0,0) \\
&= {\Big(\frac{1}{2}\Big)}^2\Big[P(N_n<\tau)+P(N_p+N_n<\tau)\Big] 
\end{split}\nonumber
\end{align}
\begin{align}\begin{split}
&\overset{(a)}= {\Big(\frac{1}{2}\Big)}^2\Bigg[1-Q\left(\frac{\tau}{\sigma}\right)\\
&+1-Q\left(\frac{\tau-n(p_2-p_1)}{\sqrt{n[p_2(1-p_2)+p_1(1-p_1)]+\sigma^2}}\right) \Bigg],
\end{split}\label{p(0,0)} \\
\begin{split}
&P_a(0,1) = P_b(0,0,1) + P_b(1,0,1)\\
&= {\Big(\frac{1}{2}\Big)}^2\Big[P(N_n\geq\tau)+P(N_p+N_n\geq\tau)\Big] \\
&={\Big(\frac{1}{2}\Big)}^2\Bigg[Q\left(\frac{\tau}{\sigma}\right)+Q\left(\frac{\tau-n(p_2-p_1)}{\sqrt{n[p_2(1-p_2)+p_1(1-p_1)]+\sigma^2}}\right) \Bigg],
\end{split} \nonumber \\
\begin{split}
&P_a(1,0) = P_b(0,1,0) + P_b(1,1,0) \\
&= {\Big(\frac{1}{2}\Big)}^2\Big[P(N_c+N_n< \tau)+P(N_p+N_c+N_n< \tau)\Big] \\
&= {\Big(\frac{1}{2}\Big)}^2\Bigg[1-Q\left(\frac{\tau-n p_1}{\sqrt{n p_1(1-p_1)+\sigma^2}}\right)\\
&+1-Q\left(\frac{\tau-n p_2}{\sqrt{n[p_2(1-p_2)+2p_1(1-p_1)]+\sigma^2}}\right) \Bigg], \\
&P_a(1,1) = P_b(1,1,1) + P_b(0,1,1) \\
&= {\Big(\frac{1}{2}\Big)}^2\Big[P(N_p+N_c+N_n\geq\tau)+P(N_c+N_n\geq\tau)\Big] \\
&= {\Big(\frac{1}{2}\Big)}^2 \Bigg[ Q\left(\frac{\tau-n p_2}{\sqrt{n[p_2(1-p_2)+2p_1(1-p_1)]+\sigma^2}}\right)\\
&+Q\left(\frac{\tau-n p_1}{\sqrt{n p_1(1-p_1)+\sigma^2}}\right) \Bigg] 
\end{split} \nonumber 
\end{align}
where, $P_a(X,Y)$ indicates the probability of $X$ sent and $Y$ received, and $P_b(Z,X,Y)$ is the probability of $X$ sent, $Y$ received, and $Z$ previously sent. $Q(\cdot)$ is the tail probability of the normal distribution. In (\ref{p(0,0)}), (a) results from (\ref{normal}) and the relationship between $Q$ function and the normal distribution. 

\subsubsection{Q-ICSK}
In Q-ICSK systems, three threshold values ($\tau_1$, $\tau_2$, and $\tau _3$) are required to represent four different symbols. The probabilities, $P_a(X,Y)$, can be obtained in a similar way as in B-ICSK systems, and only $P_a(0,0)$ is shown here as an example:
\begin{align}
\begin{split}
&P_a(0,0) = P_b(0,0,0) + P_b(1,0,0) +P_b(2,0,0) + P_b(3,0,0) \\
&=  {\Big(\frac{1}{4}\Big)}^2\Big[P(N_n<\tau_1) + 3P(N_p+N_n<\tau_1)\Big]\\
&= {\Big(\frac{1}{4}\Big)}^2\Bigg[1- Q\left(\frac{\tau_1}{\sigma}\right)\\
&+3-3Q\left(\frac{\tau_1-n(p_2-p_1)}{\sqrt{n[p_2(1-p_2)+p_1(1-p_1)]+\sigma^2}}\right)\Bigg].\end{split}\nonumber\end{align}
For other expressions, please see Appendix~\ref{appendix:Q-ICSK}.

\subsection{Molecular-Type-Based}
\label{Subsec:MoSK}
When different types of molecules represent different symbols, the technique is known as molecular-type-based modulation, referred to as molecule shift keying (MoSK)~\cite{Kuran_ICC11}. Unlike the work in~\cite{Kuran_ICC11}, we use here a set of isomers. We thus name it isomer-based MoSK (IMoSK). The IMoSK system requires only one threshold to be detected by a receiver nanomachine, which makes it simpler than the CSK system. For systematic analysis, we can choose one among the several aforementioned isomer sets; a modulation order can be determined by the sets used. For example, hexoses have a modulation order of up to $32$, trioses of $4$. For simplicity, in the first we apply, as was done in~\cite{Kuran_ICC11}, the AWGN model. The mutarotation effect explained in Section~\ref{SubSec:Isomer} is also considered in the second part.
\subsubsection{B-IMoSK-AWGN}
Only AWGN considered,
\begin{align}\begin{split}
P_a(\alpha,\alpha)_{AWGN} &= P_b(\alpha,\alpha,\alpha) + P_b(\beta,\alpha,\alpha) \\
&={\Big(\frac{1}{2}\Big)}^2\Big[P(N_p+N_c+N_n\geq\tau)P(N_n<\tau)\\
&+P(N_c+N_n\geq\tau)P(N_p+N_n<\tau)\Big],\nonumber\end{split}\\
\begin{split}
P_a(\alpha,\beta)_{AWGN} &= P_b(\alpha,\alpha,\beta) + P_b(\beta,\alpha,\beta) \\
&={\Big(\frac{1}{2}\Big)}^2\Big[P(N_n\geq\tau)P(N_p+N_c+N_n<\tau)\\
&+P(N_p+N_n\geq\tau)P(N_c+N_n<\tau)\Big],\\
P_a(\beta,\alpha)_{AWGN} &= P_b(\alpha,\beta,\alpha) + P_b(\beta,\beta,\alpha) \\
&={\Big(\frac{1}{2}\Big)}^2\Big[P(N_p+N_n\geq\tau)P(N_c+N_n<\tau)\\
&+P(N_n\geq\tau)P(N_p+N_c+N_n<\tau)\Big],\nonumber\end{split} \\ \begin{split}
P_a(\beta,\beta)_{AWGN} &= P_b(\alpha,\beta,\beta) + P_b(\beta,\beta,\beta) \\
&={\Big(\frac{1}{2}\Big)}^2\Big[P(N_c+N_n\geq\tau)P(N_p+N_n<\tau)\\
&+P(N_p+N_c+N_n\geq\tau)P(N_n<\tau)\Big].\nonumber\end{split}
\nonumber\end{align}

\begin{table*}[t]
\caption{Comparisons of Modulation Techniques.}
\begin{center}
\begin{tabular}{|c|c|c|c|c|}
\hline
 & Rate & Tx Complexity & Rx Complexity & Distortion Sensitivity\\
\hline\hline
IMoSK & good & multiple kinds of messenger molecules & multiple receptors & robust\\
\hline
ICSK & moderate & one kind of messenger molecules &one receptor & highly sensitive\\ 
\hline
IRSK & moderate & two kinds of messenger molecules & two receptors & less sensitive\\
\hline
\end{tabular}
\end{center}
\label{comparison}
\end{table*}
\subsubsection{B-IMoSK-Mutarotation Considered ($\alpha$-$D$-glucopyranose $\Leftrightarrow$ $\beta$-$D$-glucopyranose)}
\label{subsub:muta}
When $\alpha$- and $\beta$- $D$-glucopyranose are chosen for the system, possibilities arise, due to mutarotation effect, of incorrect decoding, such as $\alpha$- sent, $\beta$- received, or  $\beta$- sent, $\alpha$- received. Thus, we propose using the following expressions to calculate error probabilities, $P_{\alpha}(\alpha, \beta)$ and $P_{\alpha}(\beta, \alpha)$, considering the mutarotation process. The $\alpha$- or $\beta$- form sent varies its number with time\cite{muta}, and the number can be calculated by observing the specific optical rotation. $R_{t\alpha}$ or $R_{t\beta}$ (observed specific optical rotation at time $t$) minus $R_{eq}$ (at equilibrium) divided by $R_{\alpha}$ or $R_{\beta}$ (at time $0$) minus $R_{eq}$ has a linear relationship with time ($T_s$) at about 36.5$^{\circ}C$, body temperature. $R_{t\alpha}$ and $R_{t\beta}$ values are calculated by $R_{\alpha}$ and $R_{\beta}$, and $R_{eq}$ can be found in \cite{muta}. The number of $\alpha$- and $\beta$ forms existing after time $T_s$ is obtained as shown below. If the number of $\beta$- form exceeds the threshold value after $T_s$ when $\alpha$- sent, $\frac{\beta}{n}$ value is added to the error term. 
Thus, $P_a (\alpha, \beta)$ can be calculated as
\begin{align}
&\frac{R_{t\alpha}-R_{eq}}{R_\alpha-R_{eq}}=-\frac{0.99}{3600}t_s+1,   \nonumber\\
&R_{t\alpha}=\frac{{n_{\alpha}}R_{\alpha}+{n_{\beta}}R_{\beta}}{n_{\alpha}+n_{\beta}}=\frac{{n_{\alpha}}R_{\alpha}+(n-n_{\alpha})R_{\beta}}{n},  \nonumber \\ 
&n_{\alpha}=\frac{(R_{t\alpha}-R_{\beta})n}{R_{\alpha}-R_{\beta}}, ~n_{\beta}=n-n_{\alpha}.   \nonumber\\
&\text{If } n_{\beta}\geq\tau,~P_a(\alpha,\beta)=P_a(\alpha,\beta)_{AWGN}+\frac{n_{\beta}}{n}. \nonumber\end{align}
Also, $P_a(\beta,\alpha)$ can be calculated as
\begin {align}
&\frac{R_{eq}-R_{t\beta}}{R_{eq}-R_{\beta}}=-\frac{0.99}{3600}t_s+1, \nonumber\\
&R_{t\beta}=\frac{{n_{\alpha}}R_\alpha+{n_{\beta}}R_\beta}{n_{\alpha}+n_{\beta}}=\frac{(n-n_{\beta})R_\alpha+{n_{\beta}}R_\beta}{n},\nonumber\\
&n_{\beta}=\frac{(R_{t\beta}-R_\alpha)n}{R_\beta-R_\alpha}, ~n_{\alpha}=n-n_{\beta}.\nonumber  \\
&\text{If } n_{\alpha}\geq\tau, \,P_a(\beta,\alpha)=P_a(\beta,\alpha)_{AWGN}+\frac{n_{\alpha}}{n} \nonumber \end{align} 
where, $R_{eq}$ = 52.7$^\circ$, $R_{\alpha}$=112.2$^\circ$, and $R_{\beta}$=18.7$^\circ$~\cite{muta}. The parameters $n_{\alpha}$ and $n_{\beta}$ indicate the number of $\alpha$- form and $\beta$- form molecules, respectively, and $n$ is the number of total molecules transmitted. Optical specific rotation of the chemical compound is defined as the observed angle of optical rotation when plane-polarized light passes through the solution (i.e., $D$-glucopyranoses).
\footnote{It could be measured by a polarimeter, and there is a linear relationship between the observed rotation and the concentration of the compound~\cite{OCHEM}.}

\subsubsection{32-IMoSK}
When using hexoses, the system has a maximum modulation order of $32$. Probabilities of $X$ sent and $Y$ received for all $X$ and $Y$ values are obtained similarly.
\begin{align}\begin{split}
\text{If $X$=$Y$}\text{ (i.e.,}&\text{ if $X$ sent, $X$ received.)},\\
P_a(X,Y)&=\sum_{p=1}^{32}P_b(Z,X,Y)\\
&={\Big(\frac{1}{32}\Big)}^2\Bigg[P(N_p+N_c+N_n\geq\tau)P(N_n<\tau)\\
&+31\Big(P(N_c+N_n\geq\tau)P(N_p+N_n<\tau)\Big)\Bigg].\\
\text{If $X$}{\neq}\text{$Y$} \text{ (i.e.,}&\text{ if $X$ sent, $Y$ received)},\\
P_a(X,Y)&=\sum_{p=1}^{32}P_b(Z,X,Y)
\end{split} \nonumber 
\end{align}
\begin{align}\begin{split}
&={\Big(\frac{1}{32}\Big)}^2\Bigg[P(N_p+N_n\geq\tau)P(N_c+N_n<\tau)\\
&+31\Big(P(N_n\geq\tau)P(N_p+N_c+N_n<\tau)\Big)\Bigg].
\end{split}
\nonumber\end{align}

\subsection{Molecular-Ratio-Based}
\label{Subsec:RSK}
In addition to the concentration-based and the molecular-type-based modulations, this paper also suggests a new modulation technique as ratio-based modulation that can be referred to as isomer-based ratio shift keying (IRSK). IRSK encodes the information based on the ratios of messenger molecules. Deploying this technique yields several benefits. First, it can have a high, theoretically infinite, modulation order. Moreover, we need only two types of molecules in the simplest system though it is possible to deploy many types of molecules for more complex systems. Even though it seems quite similar to the ICSK technique, the ratio-based technique can be simpler at the detection site since only ratios, not the concentration, need be detected. It is also more robust than the ICSK system when a channel is distorted (it will be specified in Section~\ref{Subsec: comparison}).  As can be seen in Section~\ref{Subsec: Channel}, the number of messenger molecules from the current and the previous symbols can be represented as a binomial distribution, and eventually, a normal distribution. Since we use two types of molecules in one symbol duration for a simple IRSK system, the notations can be modified as follows:
\begin{align}
&N_{c1}  \sim \mathcal{N}(np_{11}, np_{11}(1-p_{11})),\nonumber\\
&N_{p1} \sim \mathcal{N}(np_{12},np_{12}(1-p_{12}))-\mathcal{N}(np_{11},np_{11}(1-p_{11})),\nonumber\\
&N_{c2}  \sim \mathcal{N}(np_{21}, np_{21}(1-p_{21})),\nonumber\\
&N_{p2} \sim \mathcal{N}np_{22},np_{22}(1-p_{22}))-\mathcal{N}(np_{21},np_{21}(1-p_{21}))
\nonumber  \end{align}
where, $N_{c1}$ and $N_{p1}$ are the number of the first molecule from the current and the previous symbols received in the current symbol duration, respectively, and $N_{c1}$ and $N_{p1}$ are of the second molecule. Also, $p_{ij}$ represents the hitting probability of $i$th molecule for $j$$T_s$.

There is one tip for picking up the proper kinds of messenger molecules to deploy the IRSK system. As mentioned in Section~\ref{subsub:muta}, mutarotation can, during propagation, occur in pairs of $\alpha$ and $\beta$ forms. Thus, even though the effect is not significant, it is better not to choose the pair as the messenger molecules. A binary IRSK (B-IRSK) system is the same as a B-MoSK in that two symbols can be represented as the ratio 1:0, or 0:1 of two types of molecules. Thus, we here analyze a Quadrature-IRSK (Q-IRSK) system, which is addressed in Appendix \ref{appendix:Q-IRSK}, and compare the result with those of ICSK and IMoSK systems.

\subsection{Comparisons}
\label{Subsec: comparison}
The three techniques in Sections~\ref{Subsec:CSK}, \ref{Subsec:MoSK}, and \ref{Subsec:RSK} have pros and cons, and one must be chosen depending on several conditions. The comparisons are summarized here. First, achievable rates can be better for IMoSK systems unless it is binary case. This is because ICSK and IRSK systems have more thresholds, which means shorter minimum distance (i.e., concentration difference) between symbols. The IMoSK system, however, has to generate and detect multiple kinds of messenger molecules whereas the ICSK and IRSK systems need generate only one or two. Thus, transmitter and receiver complexities are higher in IMoSK systems. Lastly, channel condition should also be considered to apply the proper modulation technique. That is, when the channel is distorted by any factor, the three systems are affected differently. In the worst case, there might exist some materials in the medium (i.e., channel) influencing the number of molecules named $a$. Then, the IMoSK system deploying molecules $a$, $b$, $c$, and $d$ has an additional error probability of $\frac{1}{4}$, ICSK deploying only molecule $a$ has an additional error probability of $1$, and IRSK deploying the molecules $a$ and $b$ has the error probability of $\frac{1}{2}$. 

\begin{table}[!t]
\caption{Simulation Parameters.}
\begin{center}
\begin{tabular}{|c|c|}
\hline
Parameters & Values \\
\hline\hline
$T_s$~\cite{Kuran_NCN10}& 5.9$sec$ \\ \hline
$d$~\cite{Kuran_ICC11}& 16 $\mu$m\\ \hline
$P_\text{hit}$ for $T_s$ & 0.6097 \\ \hline
$P_\text{hit}$ for $2T_s$ & 0.7208 \\ \hline
Radius of the hexoses~\cite{hexose_radi} & 0.38 $nm$ \\ \hline
$D$ of hexoses (\ref{D}) &597.25 ${\mu m}^2/sec$ \\ \hline
$\triangle H_{\text{hexose}}$ & 1271 $kJ/mol$\\ \hline
Viscosity of the water & 0.001$kg/sec \cdot m$\\ \hline
Temperature (body temperature) & 36.5 $^{\circ}C$ = 310$K$ \\ \hline
\end{tabular}
\end{center}
\label{parameters}
\end{table} 

\begin{figure}[!t]
 \centerline{\resizebox{1\columnwidth}{!}{\includegraphics{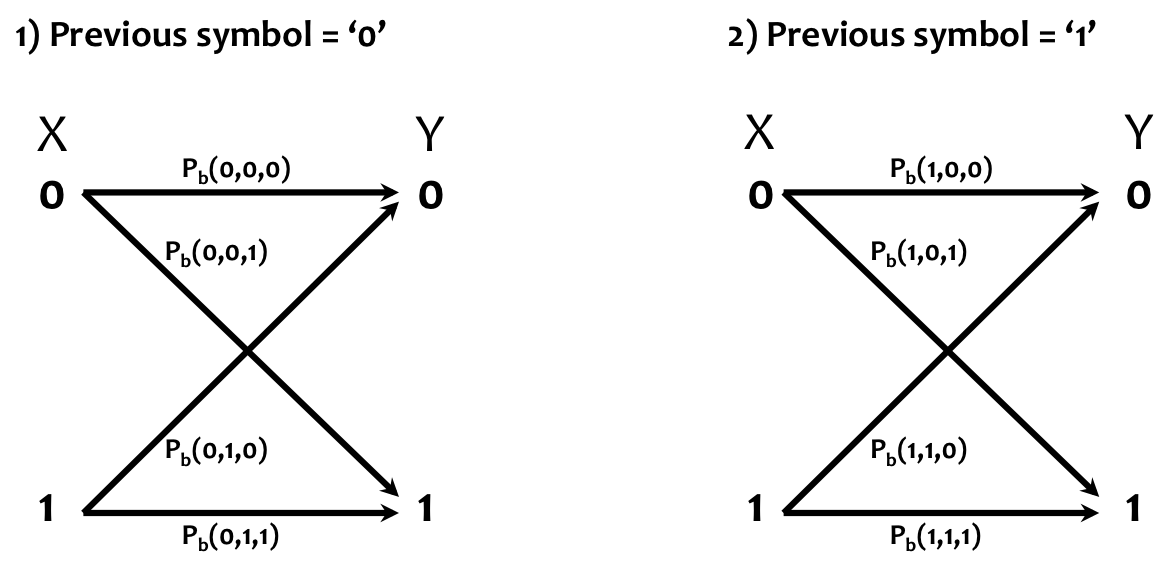}}}
  \caption{Binary symmetric channels of the binary channel model depending on the previous symbol ($Z$). The $X$ and $Y$ denote the transmitted and the received symbol, respectively, and $P_b(Z,X,Y)$ is the probability of $Z$ previously sent, $X$ currently sent, and $Y$ received.}
  \label{Fig:BSC}
\end{figure}

As specified above, each system has its advantages and disadvantages, and the most appropriate technique can be found given different situations. It will be numerically proven in Section~\ref{Sec:Num}.
Please see Table~\ref{comparison} for the summary.

\begin{figure}[!t]
 \centerline{\resizebox{1\columnwidth}{!}{\includegraphics{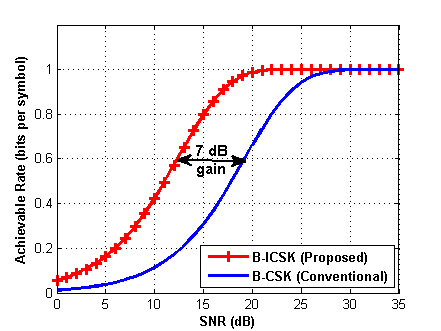}}}
  \caption{Achievable rate comparisons of the conventional insulin based CSK system and the proposed ICSK system using one kind of aldohexose isomers.}
  \label{Fig:CSK_comparison}
\end{figure}

\begin{figure}[!t]
 \centerline{\resizebox{1\columnwidth}{!}{\includegraphics{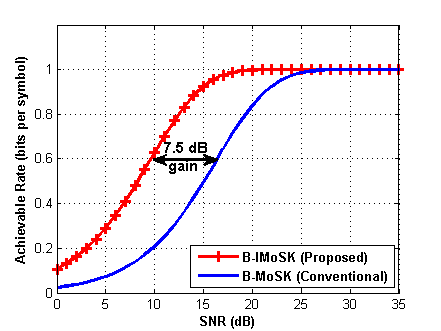}}}
  \caption{Achievable rate comparisons of the conventional insulin based MoSK system and the proposed IMoSK system using two kinds of aldohexose isomers.}
  \label{Fig:MoSK_comparison}
\end{figure}

\section{Numerical Results}
\label{Sec:Num}
Assume that all the hexoses have the same physical properties: size, diffusion coefficient, and the enthalpy of formation.
$P_\text{hit}$ value is calculated by the same numerical calculation used in \cite{Kuran_NCN10}, and it is approximated for the hexoses by assuming it to be proportional to the diffusion coefficient. Therefore, we have the hitting probabilities shown in Table~\ref{parameters} and apply these to the proposed techniques explained in Section~\ref{Sec:ModTech}.
Here, we define the achievable rate $R$ that maximizes the mutual information $I(X;Y)$ as follows:
\begin{align}\begin{split}
&I(X;Y)=\sum_X \sum_Y P_a(X,Y)\log_2{\frac{P_a(X,Y)}{P(X)P(Y)}}, \\
&R =\max_{\tau} I(X;Y)  
\end{split} \label{mi}
\end{align}
where, $P(X)$ and $P(Y)$ are the probabilities of events $X$ (the transmitted symbol) and $Y$ (the received symbol).

\begin{figure}[!t]
 \centerline{\resizebox{1\columnwidth}{!}{\includegraphics{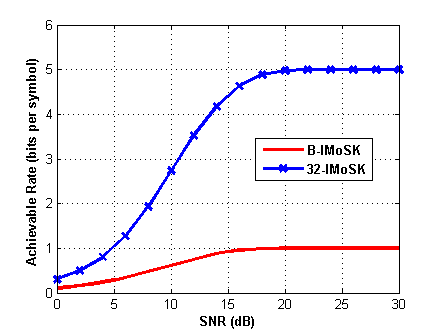}}}
  \caption{Achievable rate comparisons of 32-IMoSK and B-IMoSK. The 32 kinds of isomers of aldohexose used for 32-IMoSK, and two kinds for B-IMoSK.}
  \label{Fig:32IMoSK}
\end{figure}

\begin{figure}[!t]
 \centerline{\resizebox{1\columnwidth}{!}{\includegraphics{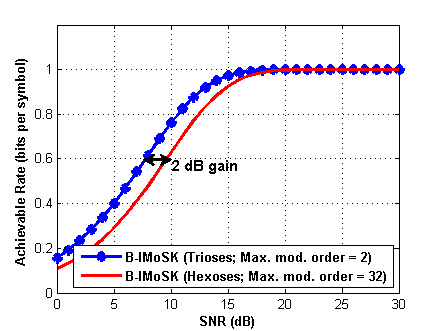}}}
  \caption{Achievable rate comparisons of the B-IMoSK system deploying trioses and hexose.}
  \label{Fig:BMoSK_tri_hex}
\end{figure}

From~(\ref{mi}), we obtain the achievable rates under various signal-to-noise ratio (SNR) values. The SNR, we use here, is defined as the ratio of the received signal power/energy to the noise power/energy. The signal power is calculated from the number of messenger molecules transmitted, the hitting probability, and the energy model derived in Section~\ref{Sec:Main}. Thus, the received number of messenger molecules is calculated using the transmitted number of molecules and the hitting probability, and the number is converted to power/energy unit by energy model. For binary channel analysis, binary symmetric channel (BSC) is used as shown in Fig.~\ref{Fig:BSC}.

Figs.~\ref{Fig:CSK_comparison} and \ref{Fig:MoSK_comparison} compare the achievable rates of the proposed method using hexoses (i.e., ICSK and IMoSK) with the conventional insulin-based method (i.e., CSK and MoSK) in binary systems. It is remarkable that, for each system, we obtain about 7~dB and 7.5~dB of SNR gain. This is due mainly to the size of the proposed messenger molecules being much smaller than that of insulin. Hence, the transmit energy of the proposed method is much less than that of the conventional method. Also, the diffusion coefficient is much larger for hexoses, which means faster propagation.

In addition, the B-IMoSK system shows a better data rate performance than B-ICSK. Fig.~\ref{Fig:32IMoSK} shows the achievable rates of 32-IMoSK and B-IMoSK using hexoses as messenger molecules. As can be seen from Fig.~\ref{Fig:32IMoSK}, the 32-IMoSK system has a maximum data rate of 5 (bits per symbol), and that of the B-IMoSK is 1 (bit per symbol). 
In Fig.~\ref{Fig:BMoSK_tri_hex}, we compare the achievable rates using trioses and hexoses. Obviously, trioses achieve higher SNR gain than hexoses due to their smaller sizes. Trioses, however, have the data transmission limit of one bit per symbol. From this result, we can conclude that trioses can be selected for a low data rate system with a higher transmission reliability and hexoses for a high data rate system. 

\begin{figure}[!t]
 \centerline{\resizebox{1\columnwidth}{!}{\includegraphics{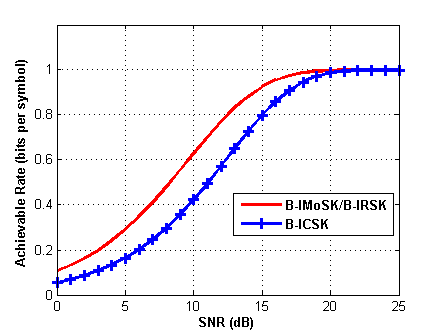}}}
  \caption{Achievable rate of binary systems using hexoses as messenger molecules.}
  \label{Fig:B_comparison}
\end{figure}

\begin{figure}[!t]
 \centerline{\resizebox{1\columnwidth}{!}{\includegraphics{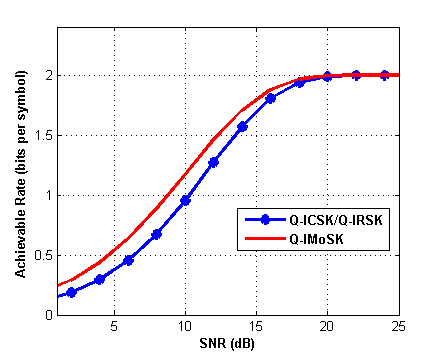}}}
  \caption{Achievable rate of Quadrature systems using hexoses as messenger molecules.}
  \label{Fig:Q_comparison}
\end{figure}

\begin{figure}[!t]
 \centerline{\resizebox{1\columnwidth}{!}{\includegraphics{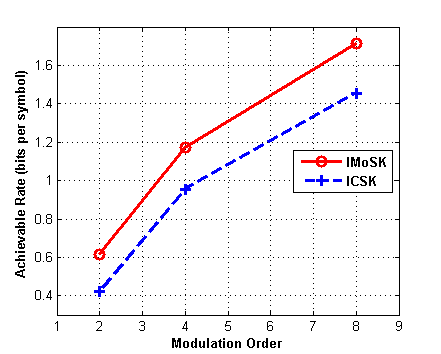}}}
  \caption{Achievable rate of IMoSK and ICSK systems in terms of modulation order when SNR is 10~dB.}
  \label{Fig:order}
\end{figure}

Figs.~\ref{Fig:B_comparison} and \ref{Fig:Q_comparison} compare the achievable rates of ICSK, IMoSK, and IRSK systems. In the binary case, the IRSK system is conceptually the same as the IMoSK while the rate goes the same way with the ICSK in the quadrature system. Fig.~\ref{Fig:order} especially shows the comparison in terms of modulation orders given SNR of 10 dB. The rate is not only higher in the IMoSK system, but the gap also increases slightly with modulation order. This can be caused by decreasing values between thresholds as modulation order increases.


\section{Conclusions}
\label{Sec:Conc}
To make nano communication feasible in practice, this work proposed novel modulation techniques using isomers as messenger molecules. We first introduced energy and channel models for our system. Next, several modulation methods were proposed so as to support up to five bits per symbol. We also compared the achievable rate performance with existing modulation methods (concentration-based and molecular-type-based), and suggested another modulation technique (ratio-based). This work differs from prior work in that it proposes practical messenger molecules and provides guidelines for selecting from among several possible candidates. For the future work, we will consider deploying multiple sets of messenger molecules and collisions among them. Also, a velocity model as well as diffusion can be added to further increase the data transmission rate.

\section*{Acknowledgment}
The authors would like to thank Dr. Songkuk Kim for helpful discussions about ratio-based modulation using isomers.

\begin{appendix}
\subsection{Q-ICSK}
\label{appendix:Q-ICSK}
Some probabilities, $P_a(X,Y)$, for Q-ICSK systems are given as follows:
\begin{align}
\begin{split}
&P_a(0,1) = P_b(0,0,1) + P_b(1,0,1) +P_b(2,0,1) + P_b(3,0,1)\\
&={\Big(\frac{1}{4}\Big)}^2\Big[P(\tau_1\leq N_n<\tau_2)+3P(\tau_1\leq N_p+N_n<\tau_2)\Big]\\
&={\Big(\frac{1}{4}\Big)}^2\Bigg[Q\left(\frac{\tau_1}{\sigma}\right)+3Q\left(\frac{\tau_1-n(p_2-p_1)}{\sqrt{n[p_2(1-p_2)+p_1(1-p_1)]+\sigma^2}}\right)\\
&-Q\left(\frac{\tau_2}{\sigma}\right)-3Q\left(\frac{\tau_2-n(p_2-p_1)}{\sqrt{n[p_2(1-p_2)+p_1(1-p_1)]+\sigma^2}}\right)\Bigg],
\end{split}\nonumber
\end{align}
\begin{align}
\begin{split}
&P_a(0,2) = P_b(0,0,2) + P_b(1,0,2) +P_b(2,0,2) + P_b(3,0,2)\\
&={\Big(\frac{1}{4}\Big)}^2\Big[P(\tau_2\leq N_n<\tau_3)+3P(\tau_2\leq N_p+N_n<\tau_3)\Big]\\
&={\Big(\frac{1}{4}\Big)}^2\Bigg[Q\left(\frac{\tau_2}{\sigma}\right)+3Q\left(\frac{\tau_2-n(p_2-p_1)}{\sqrt{n[p_2(1-p_2)+p_1(1-p_1)]+\sigma^2}}\right)\\
&-Q\left(\frac{\tau_3}{\sigma}\right)-3Q\left(\frac{\tau_3-n(p_2-p_1)}{\sqrt{n[p_2(1-p_2)+p_1(1-p_1)]+\sigma^2}}\right)\Bigg],
\end{split}\nonumber
\end{align}
\begin{align}
\begin{split}
&P_a(0,3) = P_b(0,0,3) + P_b(1,0,3) +P_b(2,0,3) + P_b(3,0,3)\\
&= {\Big(\frac{1}{4}\Big)}^2\Big[P(N_n\geq\tau_3) + P(N_p+N_n\geq\tau_3)\Big]\\
&= {\Big(\frac{1}{4}\Big)}^2\Bigg[Q\left(\frac{\tau_3}{\sigma}\right)+ 3Q\left(\frac{\tau_3-n(p_2-p_1)}{\sqrt{n[p_2(1-p_2)+p_1(1-p_1)]+\sigma^2}}\right)\Bigg],
\end{split}\nonumber
\end{align}
\begin{align}
\begin{split}
&P_a(1,0) = P_b(0,1,0) + P_b(1,1,0) +P_b(2,1,0) + P_b(3,1,0)\\
&={\Big(\frac{1}{4}\Big)}^2\Big[P(N_c+N_n<\tau_1)+3P(N_p+N_c+N_n<\tau_1)\Big]\\
&={\Big(\frac{1}{4}\Big)}^2\Bigg[1- Q\left(\frac{\tau_1-np_1}{\sqrt{np_1(1-p_1)+\sigma^2}}\right)\\
&+3-3Q\left(\frac{\tau_1-np_2}{\sqrt{n[p_2(1-p_2)+2p_1(1-p_1)]+\sigma^2}}\right)\Bigg],
\end{split} \nonumber
\end{align}
\begin{align}
 \begin{split}
&P_a(1,1) = P_b(0,1,1) + P_b(1,1,1) +P_b(2,1,1) + P_b(3,1,1) \\
&={\Big(\frac{1}{4}\Big)}^2\Big[P(\tau_1\leq N_c+N_n<\tau_2)+3P(\tau_1\leq N_p+N_c+N_n<\tau_2)\Big]\\
&={\Big(\frac{1}{4}\Big)}^2\Bigg[Q\left(\frac{\tau_1-n(p_1)}{\sqrt{n p_1(1-p_1)+\sigma^2}}\right)-Q\left(\frac{\tau_2-n p_1}{\sqrt{n p_1(1-p_1)+\sigma^2}}\right)\\
&+3\Bigg(Q\left(\frac{\tau_1-n p_2}{\sqrt{n[p_2(1-p_2)+2p_1(1-p_1)]+\sigma^2}}\right)\\
&-Q\left(\frac{\tau_2-np_2}{\sqrt{n[p_2(1-p_2)+2p_1(1-p_1)]+\sigma^2}}\right) \Bigg)  \Bigg], 
\end{split}\nonumber
\end{align}
\begin{align}
\begin{split}
&P_a(1,2) = P_b(0,1,2) + P_b(1,1,2) +P_b(2,1,2) + P_b(3,1,2)\\
&={\Big(\frac{1}{4}\Big)}^2\Big[P(\tau_2\leq N_c+N_n<\tau_3)+3P(\tau_2\leq N_p+N_c+N_n<\tau_3)\Big]\\
&={\Big(\frac{1}{4}\Big)}^2\Bigg[Q\left(\frac{\tau_2-np_1}{\sqrt{np_1(1-p_1)+\sigma^2}}\right)-Q\left(\frac{\tau_3-np_1}{\sqrt{np_1(1-p_1)+\sigma^2}}\right)\\
&+3\Bigg(Q\left(\frac{\tau_2-np_2}{\sqrt{n[p_2(1-p_2)+2p_1(1-p_1)]+\sigma^2}}\right)\\
&-Q\left(\frac{\tau_3-np_2}{\sqrt{n[p_2(1-p_2)+2p_1(1-p_1)]+\sigma^2}}\right)\Bigg)  \Bigg],
\end{split}\nonumber
\end{align}

\begin{align}
\begin{split}
&P_a(1,3) = P_b(0,1,3) + P_b(1,1,3) +P_b(2,1,3) + P_b(3,1,3)\\
&={\Big(\frac{1}{4}\Big)}^2\Big[P(N_c+N_n\geq\tau_3)+3P(N_p+N_c+N_n\geq\tau_3)\Big]\\
&={\Big(\frac{1}{4}\Big)}^2\Bigg[Q\left(\frac{\tau_3-np_1}{\sqrt{np_1(1-p_1)+\sigma^2}}\right)\\
&+3Q\left(\frac{\tau_3-np_2}{\sqrt{n[p_2(1-p_2)+2p_1(1-p_1)]+\sigma^2}}\right)\Bigg]. 
 \end{split}\nonumber\end{align}

\subsection{Q-IRSK}
\label{appendix:Q-IRSK}
The probabilities, $P_a(s_X,s_Y)$, for Q-IRSK systems are given, and here, $s_1, s_2, s_3,$ and $s_4$ denote four different symbols (i.e., symbol set, $S=\{s_1, s_2, s_3, s_4\}$).

\begin{align}\begin{split}
P_a(s_X,s_Y)& = \Big(\frac{1}{4}\Big)^2 \sum_{s_Z \in S}P_b(s_Z,s_X,s_Y)\\
P_b(s_Z,s_X,s_Y)& = P(\tau_{Y-1} \leq \alpha_1 N_{p1} + \beta_1 N_{c1} + N_{n1} < \tau_{Y})\\
&\times P(\tau_{-Y+4} \leq \alpha_2 N_{p2} + \beta_2 N_{c2} + N_{n2} < \tau_{-Y+5})\\
\end{split}\nonumber\end{align}
where, $\tau_0$ and $\tau_4$ indicate $-\infty$ and $\infty$, respectively.

\noindent If $s_Z=s_1$, then $\alpha_1$ = 0, and $\alpha_2$ =1 .\\
If $s_Z=s_2$ or $s_3$, then $\alpha_1$ = 1, and $\alpha_2$ = 1.\\
If $s_Z=s_4$, then $\alpha_1$ = 1, and $\alpha_2$ = 0.\\
If $s_X=s_1$, then $\beta_1$ = 0, and $\beta_2$ =1 .\\
If $s_X=s_2$ or $s_3$, then $\beta_1$ = 1, and $\beta_2$ = 1.\\
If $s_X=s_4$, then $\beta_1$ = 1, and $\beta_2$ = 0.

\end{appendix}

\bibliographystyle{IEEEtran}

\bibliography{references_JSAC}

\begin{biography}{Na-Rae Kim}
(S'12) received her B.S. degree in Chemical Engineering from Yonsei University, Korea in 2011. She is  now with the School of Integrated Technology at the same university and is working toward the Ph.D. degree. She was an exchange student at University of California, Irvine in USA in 2009. 

Ms. Kim is the recipient of the travel grant from the IEEE International Conference on Communications in 2012. 
\end{biography}

\begin{biography}{Chan-Byoung Chae}
(S'06 - M'09 - SM'12) is an Assistant Professor in the School of Integrated Technology, College of Engineering, Yonsei University, Korea. He was a Member of Technical Staff (Research Scientist) at Bell Laboratories, Alcatel-Lucent, Murray Hill, NJ, USA from 2009 to 2011. Before joining Bell Laboratories, he was with the School of Engineering and Applied Sciences at Harvard University, Cambridge, MA, USA as a Post-Doctoral Research Fellow. He received the Ph. D. degree in Electrical and Computer Engineering from The University of Texas (UT), Austin, TX, USA in 2008, where he was a member of the Wireless Networking and Communications Group (WNCG). 

Prior to joining UT, he was a Research Engineer at the Telecommunications R\&D Center, Samsung Electronics, Suwon, Korea, from 2001 to 2005. He was a Visiting Scholar at the WING Lab, Aalborg University, Denmark in 2004 and at University of Minnesota, MN, USA in August 2007. While having worked at Samsung, he participated in the IEEE 802.16e standardization, where he made several contributions and filed a number of related patents from 2004 to 2005. His current research interests include capacity analysis and interference management in energy-efficient wireless mobile networks and nano (molecular) communications. He serves as an Editor for the \textsc{IEEE Trans. on Wireless Communications} and \textsc{IEEE Trans. on Smart Grid}. He is also an Area Editor for the \textsc{IEEE Jour. Selected Areas in Communications} (nano scale and molecular networking). He is an IEEE Senior Member.

Dr. Chae is the recipient of the IEEE Dan. E. Noble Fellowship in
2008, the Gold Prize (1st) in the 14th Humantech Thesis Contest,
and the KSEA-KUSCO scholarship in 2007. He also received the Korea
Government Fellowship (KOSEF) during his Ph. D. studies.
\end{biography}

\end{document}